\documentclass[onecolumn, trackchanges, twocolappendix]{aastex63}

\usepackage{graphicx}
\usepackage{apjfonts}
\usepackage{amsmath}
\usepackage[raggedright]{subfigure}
\usepackage{color}
\usepackage{hyperref}
\usepackage{acronym}
\usepackage[english]{babel} 
\usepackage {lmodern}

\shorttitle{Formation of ULPPs} 
\shortauthors{\sc Yang et al.}

\begin{document}

\title{Instability in supernova fallback disks and its effect on the formation of ultra long period pulsars}

\author[0000-0001-5532-4465]{Hao-Ran Yang}
\author[0000-0002-0584-8145]{Xiang-Dong Li}
\author[0000-0002-0822-0337]{Shi-Jie Gao}
\affiliation{School of Astronomy and Space Science, Nanjing University, Nanjing 210023, China; lixd@nju.edu.cn}
\affiliation{Key Laboratory of Modern Astronomy and Astrophysics (Nanjing University), Ministry of Education, Nanjing 210023, China}
\author[0000-0002-9739-8929]{Kun Xu}
\affiliation{School of Science, Qingdao University of Technology, Qingdao 266525, China}

\begin{abstract} 

\noindent Several pulsars with unusually long periods were discovered recently, comprising a potential population of ultra long period pulsars (ULPPs). The origin of their long periodicity is not well understood, but may be related to magnatars spun down by surrounding fallback disks. While there are few systematic investigations on the fallback disk-assisted evolution of magnetars, the instability in the disk has received little attention, which determines the lifetime of the disk. In this work we simulate the evolution of the magnetic field, spin period, and magnetic inclination angle of magnetars with a supernova fallback disk. We find that thermal viscous instability in the disk could significantly affect the formation of ULPPs. Our simulation results also reveal that a large fraction of ULPPs seem to be nearly aligned and orthogonal rotators. This might help place ULPPs above the death line in the pulse period - period derivative plane. However, some extra mechanisms seem to be required to account for radio emission of ULPPs.

\end{abstract}


\section{Introduction}
Pulsars are  magnetized rotating neutron stars (NSs) with spin periods ranging from milliseconds to several seconds. This traditional view was challenged by the discoveries of several peculiar pulsars with ultra long periods in recent years, including the 6.67 hr central compact object 1E 161348$-$5055 (hereafter 1E 1613) in the supernova remnant RCW 103 \citep{2006Sci...313..814D}, the 18.18 min, 21 min and 53.8 min radio pulsed sources GLEAM$-$X J162759.5$-$523504.3 (hereafter GLEAM$-$X J1627) \citep{2022Natur.601..526H}, GPM J1839$-$10 (hereafter GPM J1839) \citep{hurley-walker_long-period_2023} and ASKAP J193505.1+214841.0 (hereafter ASKAP J1935) \citep{2024NatAs.tmp..107C}, as well as the 76 s pulsar J0901$-$4046 (hereafter PSR J0901) \citep{2022NatAs...6..828C}. 

Because magnetic dipole radiation, even for NSs with ultra strong magnetic fields (i.e., magnetars), is unable to account for these ultra long period pulsars (ULPPs), it was proposed that they may have experienced additional spin-down torque(s). Possible scenarios include that ULLPs originate from magnetars surrounded by a supernova fallback disk  \citep{2006Sci...313..814D,2007ApJ...666L..81L}, or they are acturally white dwarf pulsars \citep{2022Ap&SS.367..108K,2022RNAAS...6...27L,2024ApJ...961..214R}. However, a constraint on the radius of the most recently published source, ASKAP J1935, ruled out an isolated magnetic white-dwarf origin \citep{2024NatAs.tmp..107C}. As to fallback disk-assisted spin-down, \citet{2000ApJ...534..373C} suggested that NSs with moderate ($\sim 10^{12}-10^{13}$ G) magnetic fields interacting with a fallback disk could explain the $\sim 5-10$ s pulse periods of anomalous X-ray pulsars (AXPs). Subsequently, numerous studies have attempted to apply this idea to investigate the evolution of young NSs with a fallback disk, aiming to account for ULPPs such as 1E 1613 \citep{2006Sci...313..814D,2007ApJ...666L..81L,2016ApJ...833..265T,2019ApJ...877..138X}, GLEAM-X J1627 \citep{2022MNRAS.513L..68G,2022ApJ...934..184R,2023ApJ...943....3T}, and GPM J1839 \citep{2023RAA....23l5018T,2024ApJ...961..214R,2024ApJ...967...24F}. While the majority of them have focused on the NS-fallback disk interaction, limited attention has been paid to the impact of the fallback disk lifetime \citep{2001ApJ...559.1032M,2001ApJ...554.1245A}, which critically depends on the condition of neutralization of the disk. It is well known that a NS can appear as a radio pulsar only when the fallback disk is destroyed or at least pushed outside the light cylinder. 
Moreover, the NS should be an oblique rotator, i.e., the spin and magnetic axes of the NS are not aligned, to allow the lighthouse effect to work. Additionally, the magnetic inclination angle $\chi$ may influence the condition for magnetospheric
particle acceleration around NSs, i.e., the so-called ``death line'' in the pulse period - period derivative plane \citep{2006ApJ...643.1139C,2012ApJ...757L..10T}. Many researchers have discussed the evolution and/or distribution of the magnetic inclination angle $\chi$ for normal radio pulsars \citep{2014MNRAS.441.1879P,2017MNRAS.466.2325A,2020MNRAS.494.3899N} and magnetars \citep{2018MNRAS.481.4169L,2020MNRAS.496.2183C}. However, the evolution of the magnetic inclination angles for NSs interacting with a fallback disk has received limited attention. 

In this work we examine whether ULPPs can be explained by magnetars with a fallback disk by simultaneously simulating the disk evolution and the NS's spin and magnetic inclination evolution. The rest of this paper is organized as follows. We first introduce the model of evolution for fallback disks and $\chi$ of magnetars in Section~\ref{s:method}. In Section~\ref{s:result}, we provided several examples to investigate how $P$ and $\chi$ evolve with initially different parameters and disk instability models, and explore the possibility for it to become an ULPP. In Section~\ref{s:observation}, we present  Monte-Carlo (MC) simulations to compare with known observations and explore the potential for detecting
radio radiation from them. Finally, the conclusions are given in Section~\ref{s:conclusion}.

\section{Method}\label{s:method}

\subsection{The evolution of a fallback disk}\label{s:fallback model}
\begin{figure}[!t]
	\centering    \includegraphics[width=.5\textwidth]{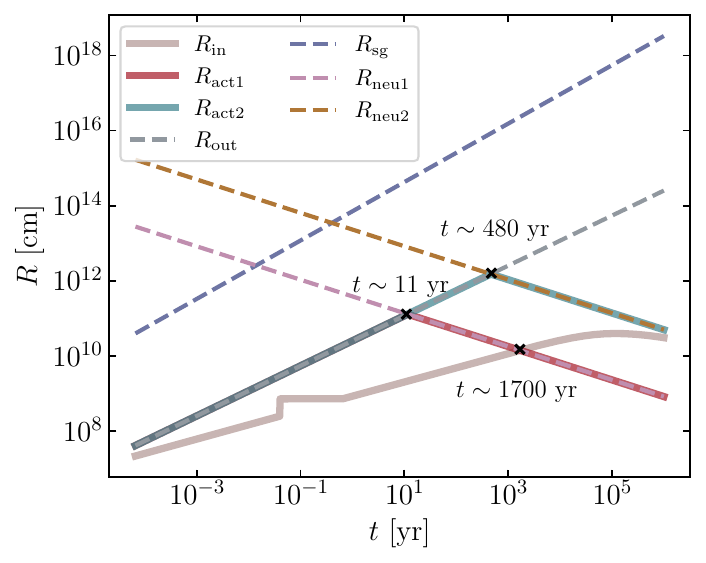}
	\caption{The evolution of several radii. The dashes lines represent several radii that may impact the evolution of outer disk, while the solid lines indicate the inner radius, and active outer radii of two models. The initial parameters are fixed to be: $M_{\rm d0}=10^{-4}\,M_{\odot}$, $R_0=10^2\rm R_{\rm s}$, $B_0=10^{15}\rm G$, $\alpha_0=45^{\circ}$, $\chi_0=60^{\circ}$ and $P_0=0.1\rm s$.}
	\label{fig:radii}
\end{figure}

Shortly after a supernova explosion, part of the ejected material could be bounded around the central compact object by its gravity, and form a fallback disk if the material possesses sufficiently large angular momentum \citep{1993ApJ...405..273W,2000ApJ...534..373C}. The evolution of a fallback disk can generally be divided into four phases, from a slim disk to a thin disk, and eventually advection-dominated accretion flow \citep[e.g.,][]{2015ApJ...814...75L}.  In this paper, we simplify the detailed process and adopt the following self-similar solutions \citep[e.g,][]{2019ApJ...877..138X}:
\begin{align}
\dot{M}&=\dot{M}_{0}(t/t_{\rm f})^{-4/3},\label{eq:mdot}\\
R_{\rm out}&=R_{\rm 0}(t/t_{\rm f})^{2/3},\label{eq:rout}\\
\rho&=\rho_{0}(\dot{M}/\dot{M}_{0}),\label{eq:rho}
\end{align}
where $\dot{M}$, $R_{\rm out}$ and $\rho$ are the mass transfer rate, the outer radius and the density of the disk, respectively,
\begin{equation}
  \dot{M}_{0}=M_{\rm d0}/t_{\rm f},\label{eq:mdot0},
\end{equation}
and
\begin{equation}
 \rho_{0}\simeq 1.53\times 10^{-4}(\dot{M}_{0}/\dot{M}_{\rm Edd,\odot})R_{\rm 0,s}^{-3/2} \rm{g\,cm^{-3}}.\label{eq:rho0}   
\end{equation}
The subscript 0 represents the values evaluated at the disk formation time $t_{\rm f}$, $R_{0,\rm s}=R_{0}/R_{\rm s}$ is the initial outer radius normalized by the Schwartzschild radius ($R_{\rm s}\simeq 4\times 10^5\,\rm cm$ for a $1.4~M_{\odot}$ NS), and $\dot{M}_{\rm Edd,\odot}(=1.43\times10^{18}\,\rm gs^{-1})$ is the Eddington limit accretion rate for a $1~M_{\odot}$ star. 
The disk formation time $t_{\rm f}$ is taken to be the initial viscous timescale $t_{\rm visc}$. \citet{2001ApJ...559.1032M} and \citet{2009ApJ...702.1309E} estimated $t_{\rm visc}$ in terms of the central temperature  of the disk and the initial disk mass, respectively. Here we adopt the maximum of them, i.e.,
\begin{equation}
    t_{\rm f}=\max (131.55 T_{\rm c,6}^{-1}R_{\rm 0,s}^{1/2}, 0.52 M_{\rm d0,-4}^{-3/7}R_{0,\rm s}^{25/14})\,\rm{s}
\end{equation}
where $T_{\rm c,6}=T_{\rm c}/10^{6}~\rm{K}$ and $M_{\rm d0,-4}=M_{\rm d0}/10^{-4}M_{\odot}$ are the normalized central temperature and initial mass of the disk, respectively\footnote{The dynamical timescale is significantly shorter than the viscous timescale in our simulations, so is not considered here.}. 
We note that the power law index in Eq.~(\ref{eq:mdot}) actually depends on opacity and disk wind \citep{1990ApJ...351...38C,  2002ApJ...565..471F,2009ApJ...700.1047C}, but it does not have significant impact on our simulation results. Thus, we fix it to be $-4/3$. 

Driven by the viscous dissipation, the disk continuously expands outward, as stated by Eq.~(\ref{eq:rout}).  Certainly, its expansion is limited by the self-gravity radius $R_{\rm sg}$ and the neutralization radius $R_{\rm neu}$. Here $R_{\rm sg}$ is the radius beyond which the self-gravity of the disk becomes important and the disk becomes gravitationally unstable \citep{1998ApJ...509...85B},
\begin{equation}
R_{\rm sg}\simeq 9.62\times 10^{10} \left(\frac{\rho}{1\ \rm{g\ cm^{-3}}}\right)^{-1/3}\ \rm{cm}. \label{eq:rsg}
\end{equation}
Gravitational instability will lead to fragmentation of the disk if the cooling time is short enough.

In addition, when the outermost disk is sufficiently cool and neutral, the material in the disk could hardly be transported along the disk, the disk becomes passive and the disk activity turns off. A lower neutralization temperature means longer lifetime of an active disk. \citet{2001ApJ...559.1032M} suggested that, similar as the accretion disks in dwarf novae and transient X-ray binaries, the fallback disks would be subject to the thermal ionization instability once the effective temperature $T_{\rm eff}$ of the outer disk drops to $\sim 6500$ K, below which the disk material becomes neutral \citep{2001NewAR..45..449L}. For an optically thick accretion disk, the relationship between $T_{\rm eff}$ and the mass flow rate is given by \citep{2002apa..book.....F}
\begin{equation}
    \sigma T_{\rm eff}^4 = \frac{3}{8\pi}\frac{GM_{*}\dot{M}}{R^3}, \label{eq:blackbody}
\end{equation}
where $\sigma$ is the Stefan-Boltzmann constant, $G$ is the gravitational constant, $R$ is the disk radius, and $M_{*}$ is the mass of the central object (NS in this paper). Combining Eqs.~(\ref{eq:mdot}) and (\ref{eq:blackbody}) we obtain the expression of the neutralization radius
\begin{equation}
    R_{\rm neu}=6.08\times10^{10}\left(\frac{T_{\rm eff}}{6500\ \rm K}\right)^{-4/3}\left(\frac{M_*}{M_{\odot}}\right)^{1/3}\left(\frac{\dot{M}_{0}}{\dot{M}_{\rm Edd,\odot}}\right)^{1/3}\left(\frac{t}{t_{\rm f}}\right)^{-4/9}\ \rm cm. \label{eq:rneu}
\end{equation}
However, \cite{2009ApJ...702.1309E} argued that the minimum temperature for an active disk could be as low as $\sim 100-300$ K, based on the work by \citet{2005ApJ...628L.155I}, who showed that the viscosities are expected to be turbulent by the magneto-rotational instability (MRI) even at temperatures $\sim 300$ K.\footnote{It is noted that MRI turbulence depends on the the ionization fraction due to strong electric fields \citep[][and references therein]{Mori_2019}. Extremely weak level of ionization can induce non-ideal MHD effects to weaken or suppress the MRI.} For typical fallback disk parameters $R_{\rm sg}$ is much larger than $R_{\rm neu}$.

Whether the disk instability results in large outbursts separated by long quiescence or rapid oscillations depends on the propagation time of heating and cooling fronts, which is given by \citep{1999MNRAS.305...79M}
\begin{equation}
t_{\rm front}=\frac{R}{\alpha c_{\rm s}}\sim \frac{h}{R}t_{\rm visc}, \label{eq:front}
\end{equation} 
where $\alpha$, $c_{\rm s}$, and $h$ are the viscosity parameter, the sound speed in the disk, and the disk height, respectively. In the standard \citet{1973A&A....24..337S} accretion disk model, $h/R\ll1$ unless for super-critical accretion.
Then heating and cooling fronts propagate on time scales much shorter than the viscous time, and the transition fronts propagate back and forth in the disk leading only to small luminosity variations \citep{2007IAUS..238..297H}. However, the disk thickness could be considerably thicker than in the standard model because of, e.g., advection-dominated flow \citep{1994ApJ...428L..13N}, amplification of the toroidal magnetic field \citep{2007MNRAS.375.1070B}, and/or X-ray irradiation of the outer disk \citep{1999MNRAS.303..139D}. In this case, local instability of an annulus may be able to lead to the global evolution of the disk, as in cataclysmic variables and transient X-ray binaries \citep[][for reviews of disk instability]{2001NewAR..45..449L,2020AdSpR..66.1004H}. To discriminate these two possibilities, we construct two models to constrain the lifetime of an active fallback disk:
\begin{enumerate}
  \item Model I -  we take the actual outer disk radius $R_{\rm act}$ of an active disk to be the minimum of $R_{\rm out}$, $R_{\rm sg}$, and $R_{\rm neu}$, which is denoted as $R_{\rm neu1}$ for $T_{\rm eff}=6500$ K (Model I-1), and $R_{\rm neu2}$ for $T_{\rm eff}=300$ K (Model I-2). An active disk exists only if $t>t_{\rm f}$ and $R_{\rm in}<R_{\rm act}$ (where the inner radius $R_{\rm in}$ of the disk is described by Eq.~[\ref{eq:rm}] in next section). This model corresponds to the case that $t_{\rm front}$ is significantly shorter than $t_{\rm visc}$.
  \item Model II - the disk becomes passive once $R_{\rm out}$ equals $R_{\rm neu1}$ (Model II-1) or $R_{\rm neu2}$ (Model II-2). This model corresponds to the case that $t_{\rm front}$ is shorter but not significantly shorter than $t_{\rm visc}$.
\end{enumerate}

Figure~\ref{fig:radii} shows the evolution of the outer disk radius with a set of initial parameters. 
It can be seen that $R_{\rm sg}$ is much larger than other radii during most of the evolution. At the time of 11 yr (480 yr), $R_{\rm out}$ intersects $R_{\rm neu1}$ ($R_{\rm neu2}$). We can expect that the disk becomes globally neutral and disrupted by the NS magnetic field after this moment in model II. In model I, the disk is assumed to keep active for longer time, until $R_{\rm neu1}$ ($R_{\rm neu2}$) intersects $R_{\rm in}$ at the time about 1700 yr ($10^6$ yr). 

\subsection{The evolution of a NS with a fallback disk}\label{s:spin model}
Before presenting the detailed evolutionary equations, we first introduce three critical radii that determine the evolutionary stages of the NS, i.e., 
\begin{align}
R_{\rm in}&\simeq 1.1\times10^9\mu_{32}^{4/7}M_{*,\odot}^{-1/7}(\dot{M}/\dot{M}_{\rm Edd,\odot})^ {-2/7}\rm{cm},\label{eq:rm}\\
R_{\rm co}&\simeq1.5\times10^{8}M_{*,\odot}^{1/3}P_{\rm s,1}^{2/3}\rm{cm},\label{eq:rco}\\
R_{\rm lc}&\simeq4.8\times10^{9}P_{\rm s,1}\rm{cm}, \label{eq:rlc}
\end{align}
where $R_{\rm in}$, $R_{\rm co}$, and $R_{\rm lc}$ are the inner radius of a disk (or the magnetospheric radius) \citep{1979ApJ...234..296G}, the co-rotation radius, and the light cylinder radius, respectively. In the above equations $\mu_{32}=\mu /10^{32}\,\rm G\,cm^3$ and $P_{\rm s,1}=P/1\rm s$ are the magnetic dipole moment and the spin period of the NS in units of $10^{32}\,\rm G\,cm^3$ and 1 s, respectively. 

A magnetized rotating NS surrounded by a fallback disk generally experience three different phases: (I) The NS is in the accretor phase if $R_{\rm in}$ is less than $R_{\rm co}$ and $R_{\rm lc}$ because of the very high mass transfer rate.  (II) When $R_{\rm in}$ becomes larger than $R_{\rm co}$ but still less than $R_{\rm lc}$, the material at the inner edge of the disk cannot be effectively accreted by the NS. Instead, it is repelled outward due to the centrifugal barrier. This phase is called the propeller phase. (III) When $R_{\rm in}$ is larger than both $R_{\rm co}$ and $R_{\rm lc}$, the disk can no longer exert any influence on the NS, and the NS enters the ejector phase, exhibiting behavior similar to that of a normal radio pulsar. 

Then, the averaged torques acting on the NS with a fallback disk are composed of three parts: the accretion torque $N_{\rm acc}$, the torques caused by magnetic field-disk interaction $N_{\rm md}$ and by magnetic dipole radiation $N_{\rm dip}$, which can be respectively expressed as \citep{2019ApJ...877..138X,2023ApJ...945....2Y}:
\begin{equation}
N_{\rm acc} =\dot{M}_{\rm acc}\left(GM_{*}R_{\rm in}\right)^{1/2},\  \mathrm{if} \ R_{\rm in}<R_{\rm co}<R_{\rm lc},\label{eq:n1}
\end{equation}
\begin{align}
N_{\rm md}=
\begin{cases}
 &-\mu^2/(3R_{\rm co}^3),\ \mathrm{if} \ R_{\rm in}<R_{\rm co}<R_{\rm lc},\\
           &-\dot{M}_{\rm in}\left(GM_{*}R_{\rm in}\right)^{1/2},\ \mathrm{if} \ R_{\rm co}\le R_{\rm in}<R_{\rm lc},\\
\end{cases}
\end{align}
\begin{equation}
N_{\rm dip} = -\mu^2/R_{\rm lc}^3 ,
\end{equation}
where $G$ is the gravitational constant, $\dot{M}_{\rm acc}$ and $\dot{M}_{\rm in}$ are the accretion rate onto the NS and the mass transfer rate at the inner edge of the disk, respectively. We assume that the accretion rates of the NS and mass transfer rate are both limited by Eddington rate unless the accretion flow is dominated by neutrino loss, that is \citep{2019ApJ...877..138X,2020ApJ...899...97E}
\begin{align}
\dot{M}_{\rm acc} =
\begin{cases}
\dot{M}_{\rm in},  & \mathrm{if} \ \dot{M}_{\rm in}>\dot{M}_{\rm cr}\ \mathrm{or}\ \dot{M}_{\rm in}<\dot{M}_{\rm Edd},\\ 
\dot{M}_{\rm Edd},  & \mathrm{else,} \\
\end{cases}\label{eq:acc}
\end{align}
and 
\begin{align}
\dot{M}_{\rm in} =
\begin{cases}
\dot{M},  & \mathrm{if} \ \dot{M}>\dot{M}_{\rm cr}\ \mathrm{or}\ \mathrm{if}\ \dot{M}<\dot{M}_{\rm cr}\ \mathrm{and}\ R_{\rm in}>R_{\rm sph};\\
\dot{M}(R_{\rm in}/R_{\rm sph}),  & \mathrm{if} \ \dot{M}<\dot{M}_{\rm cr}\ \mathrm{and}\ R_{\rm in}\le R_{\rm sph}.\\
\end{cases}\label{eq:in}
\end{align}
Here $R_{\rm sph}=3GM\dot{M}/(2L_{\rm Edd})$ and $\dot{M}_{\rm cr}\simeq 1.9\times10^{22}\rm g\ s^{-1}$ are the spherization radius, where the local accretion luminosity in the disk starts to exceed the Eddington limit \citep{1973A&A....24..337S,1999AstL...25..508L}, and the critical mass transfer rate above which neutrino loss dominates the accretion flow and all the matter transferred can accrete onto the NS \citep{1971ApJ...163..221C,1989ApJ...346..847C}. 

With that in mind, we list the differential equations for the evolution of a fallback disk-fed NS \citep{2021MNRAS.505.1775B,2023ApJ...945....2Y},
\begin{align}
    I\Omega\dot{\alpha} &= - N_{\rm acc}\sin\alpha, \label{eq:alpha} \\
    I\dot{\Omega} &= N_{\rm acc}\cos \alpha + N_{\rm md} + N_{\rm dip}(1+\sin^2\chi)-\dot{I}\Omega, \label{eq:omega}\\
    \nonumber I\Omega\dot{\chi} &= \eta A(\eta,\alpha,\chi)N_{\rm acc}\sin^2\alpha\cos\alpha \sin\chi \cos\chi+N_{\rm dip}\sin\chi \cos\chi, \label{eq:chi}
\end{align}
where $I$ is the moment of inertia of the NS, $\alpha$ is the angle between the spin axis of the NS and rotation axis of the disk, $\chi$ is the angle between the spin and magnetic axes of the NS, $\eta = 0.99$ is used to describe the accretion torque modulation within a spin period, and $A(\eta,\alpha,\chi)$ is a normalization factor  \citep[see][]{2021MNRAS.505.1775B}.

Furthermore, the magnetic field of the NS decays by Ohmic dissipation and the Hall effect in the NS’s crust. To account for this effect we adopt the following phenomenological formula \citep{2008ApJ...673L.167A}
\begin{equation}
    B(t)=B_{0}\frac{e^{-t/\tau_{\rm Ohm}}}{1+\frac{\tau_{\rm Ohm}}{\tau_{\rm Hall,0}}[1-e^{-t/\tau_{\rm Ohm}}]}, \label{eq:mag}
\end{equation}
where $B_{0}$ is the initial magnetic field of the magnetar, $\tau_{\rm Ohm}$ and $\tau_{\rm Hall,0}$ are the characteristic times for Ohmic diffusion and Hall effect, respectively.

Combining the equations for the evolution of the fallback disk and the NS, we can track how $P$, $B$, $\alpha$ and $\chi$ change with time under different initial conditions, and explore the possibility for the NS to become a pulsar. 

\section{Example Evolution}\label{s:result}
\begin{figure}[!t]
	\centering    \includegraphics[width=\textwidth]{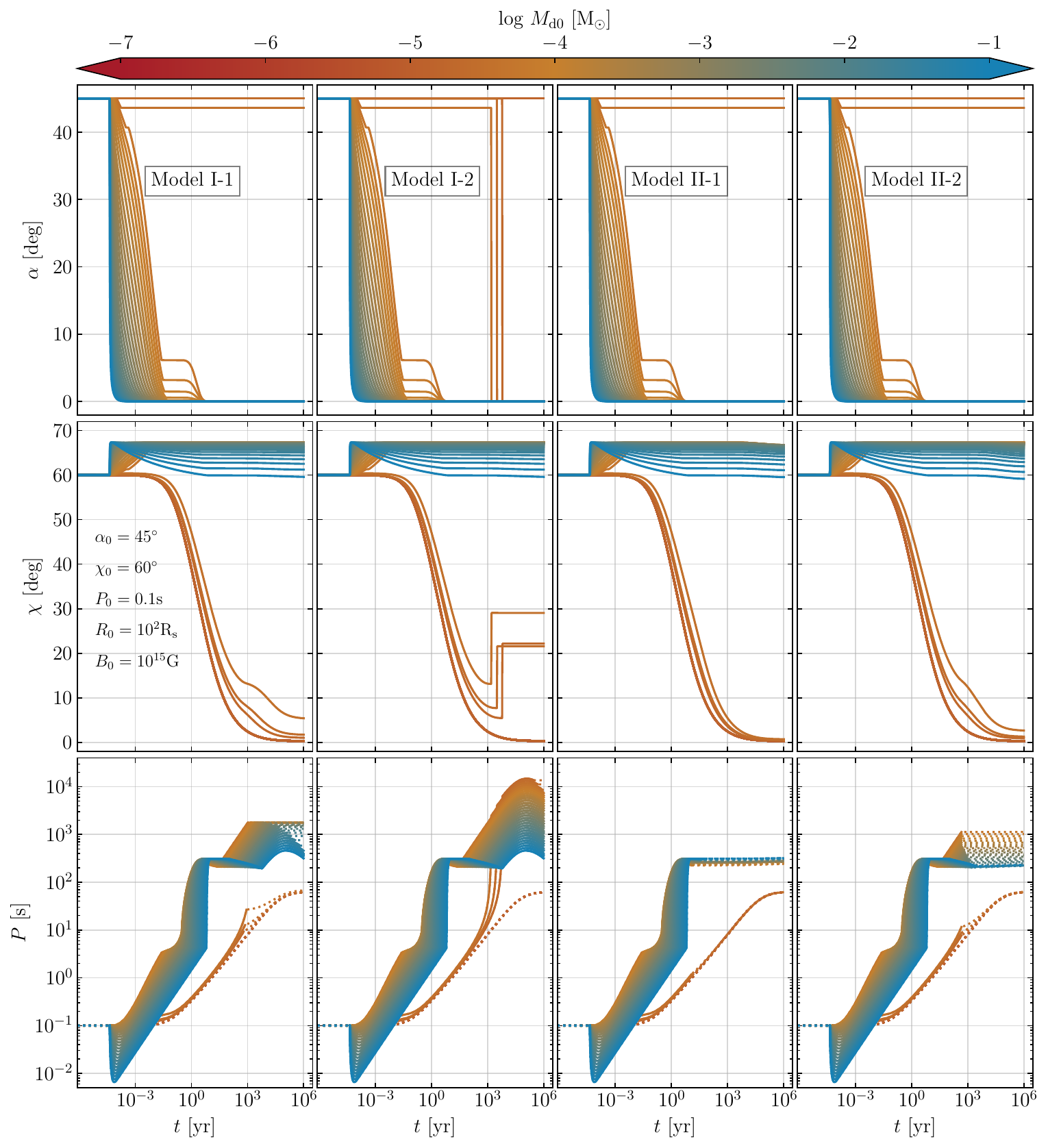}
	\caption{Evolution of $\alpha$, $\chi$, and $P$ with different initial fallback masses $M_{\rm d0}$.The four columns correspond to the results of Models I-1, I-2, II-1 and II-2. The solid and dotted parts of the lines in the bottom panels represent if the fallback disks exist or not. The other initial parameters are $R_0=10^2\rm R_{\rm s}$, $B_0=10^{15}\,\rm G$, $\alpha_0=45^{\circ}$, $\chi_0=60^{\circ}$, and $P_0=0.1\,\rm s$.}
	\label{fig:md}
\end{figure}

Figure~\ref{fig:md} shows the evolution of of $\alpha$ (upper panel), $\chi$ (middle panel), and $P$ (lower panel) in the four models with the following initial parameters:  $M_{\rm d0}=(10^{-7}-10^{-1})\,\rm M_{\odot}$, $R_0=10^2\,\rm R_{\rm s}$, $B_0=10^{15}\,\rm G$, $\alpha_0=45^{\circ}$, $\chi_0=60^{\circ}$ and $P_0=0.1\,\rm s$. We limited the evolution up to $10^6$ yr. In the lower panel the solid and dotted lines indicate an active disk and a passive disk/no disk, respectively. With low neutralization temperature the disk can persist up to $10^6$ yr, similar as in  \cite{2022ApJ...934..184R}. The corresponding evolutionary trajectories can be classified into three categories. 

Class I. If $M_{\rm d0}<10^{-5}\,\rm M_{\odot}$, $R_{\rm in}>R_{0}$ at $t=t_{\rm f}$, thus no disk is formed. The NS spins down solely by magnetic dipole radiation, which also causes $\chi$ to decrease monotonically. Note that the spin period $P$ can be spun down to $\sim 100\,\rm s$ at most even if the initial magnetic field $B_0=10^{15}\,\rm G$.

Class II. If $M_{\rm d0}\sim 2-3\times 10^{-5}\,\rm M_{\odot}$, $R_{\rm lc}<R_{\rm in}<R_{\rm out}$ at $t=t_{\rm f}$, thus the disk can form and the NS is initially in the ejector phase. The evolution of both $P$ and $\chi$ is similar as in Class I. With decreasing $\dot{M}$ and increasing $P$, $R_{\rm in}$ evolves to be less than $R_{\rm lc}$, and then less than $R_{\rm co}$. The torques exerted by the disk (i.e. $N_{\rm acc}$ and $N_{\rm md}$) begin to work, causing $\alpha$ to decrease and $\chi$ to increase \citep[see the detailed analysis in][]{2023ApJ...945....2Y}. The spin period $P$ eventually grows to $>1000\,\rm s$ driven by the magnetic field-disk interaction.

Class III. If $M_{\rm d0}\gtrsim 4\times 10^{-5}\,\rm M_{\odot}$, $R_{\rm in}<R_{\rm co}$ at $t=t_{\rm f}$, the NS resides in the accretor phase in the beginning. Both $\alpha$ and $\chi$ changes due to accretion as described in Class II. The spin period can also be spun down to thousands of seconds. The maximum period the NS gets smaller with a more massive fallback disk. 

The NS in Classes II and III can potentially evolve to be ULPPs. 


\section{Population Studies and Comparison with Observations}\label{s:observation}
\begin{figure*}[t]
	\centering \includegraphics[width=\textwidth]{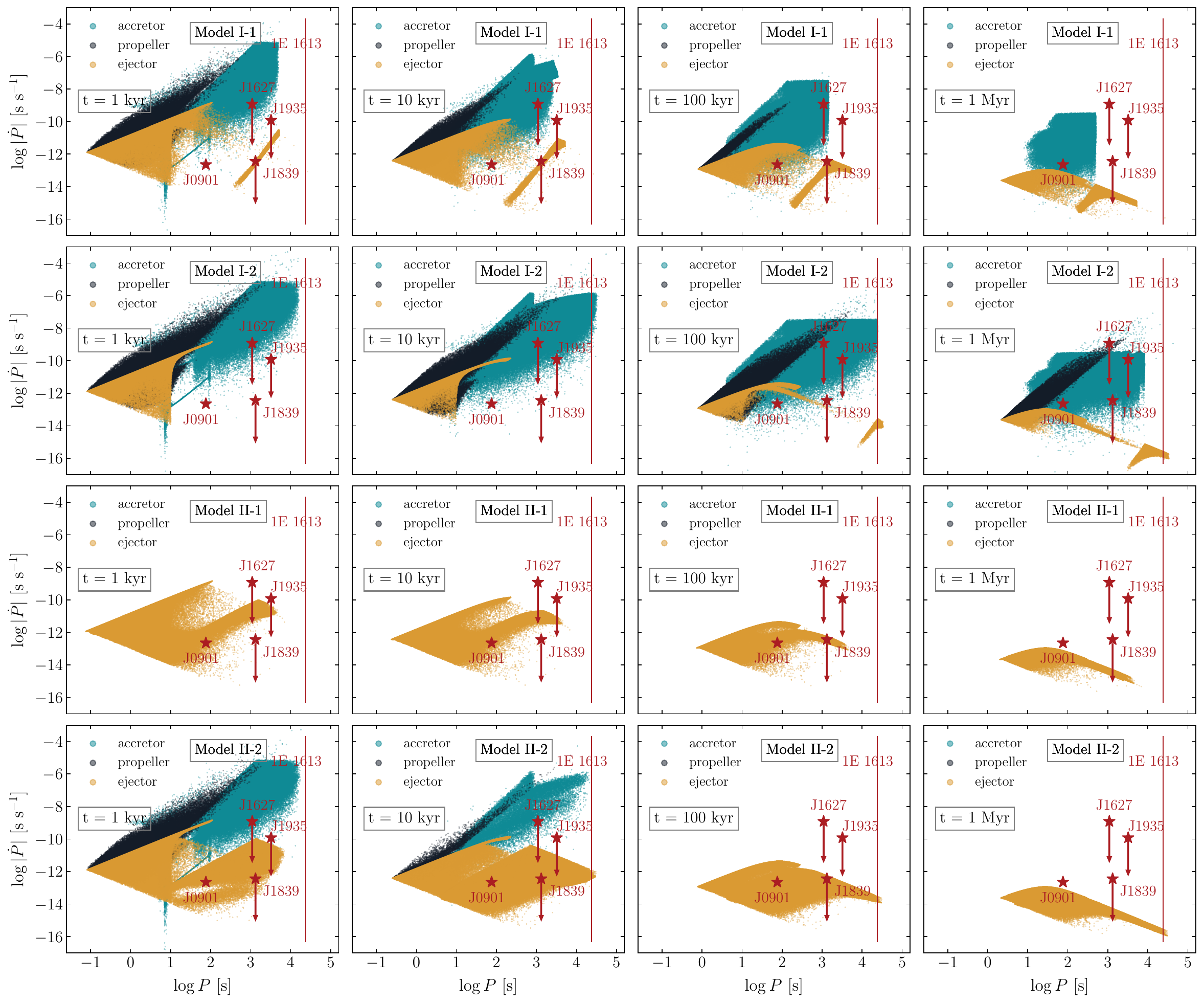}
	\caption{The period-period derivative ($P-\dot{P}$) diagrams in different models at the ages of 1, 10, 100, and $10^3$ kyr (from left to right). The red stars and vertical line represent the four ULPPs, i.e., PSR J0901-4046, GLEAM-X J162759.5-523504.3, GPM J1839-10, ASKAP J193505.1+214841.0, and the central compact object 1E 161348-5055, respectively.}
	\label{fig:mc}
\end{figure*}

\begin{figure*}[t]
	\centering \includegraphics[width=\textwidth]{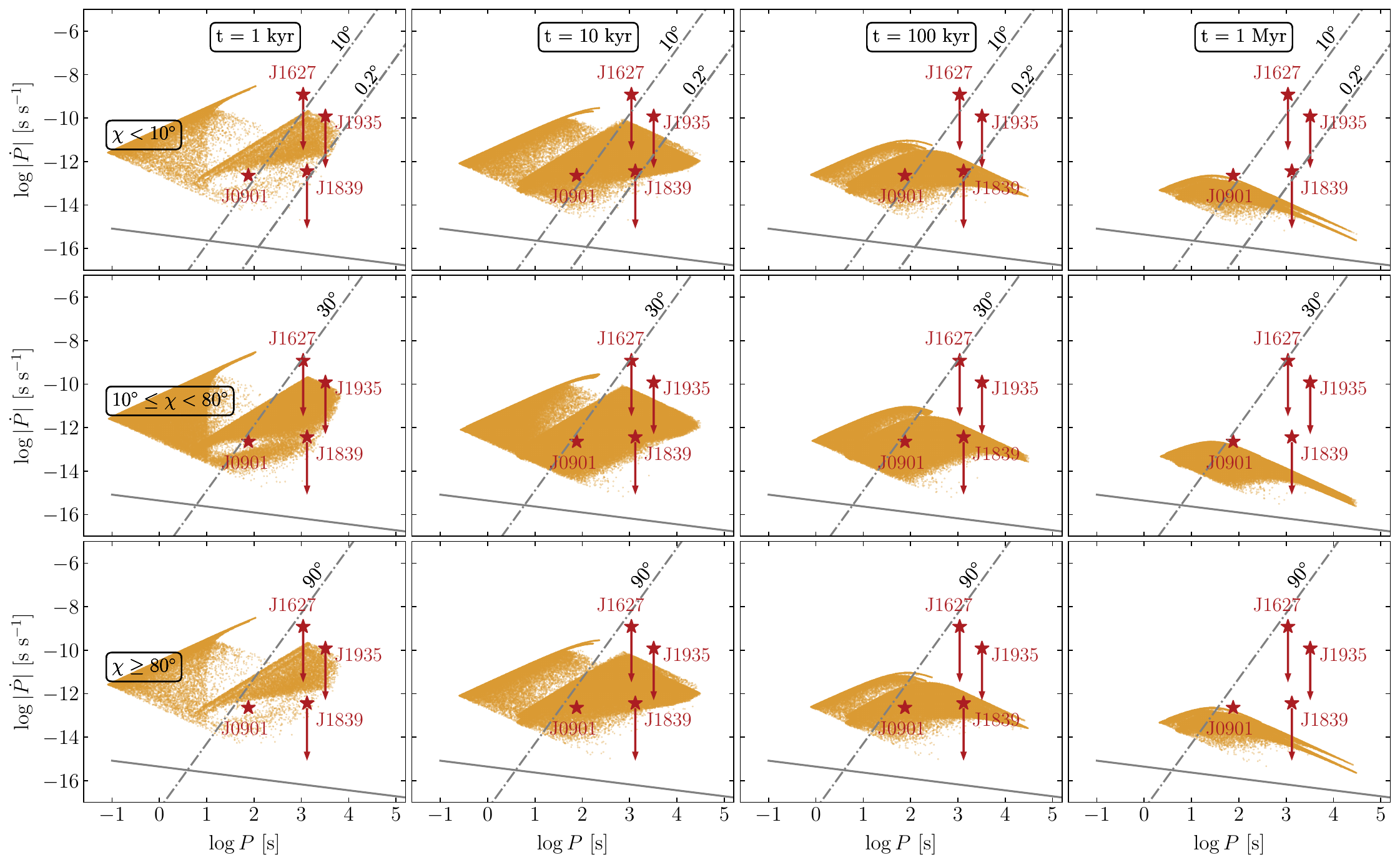}
	\caption{The distribution of ejectors in Model II-2 for different magnetic inclination angle $\chi$. The gray dot-dashed lines represent the death lines corresponding to $\chi=0.2^{\circ}$, $10^{\circ}$, $30^{\circ}$, and $90^{\circ}$, and the solid line is the modified death line from Eq.~(10) in \citet{2000ApJ...531L.135Z}.}
	\label{fig:death}
\end{figure*}

\begin{figure*}[t]
	\centering \includegraphics[width=\textwidth]{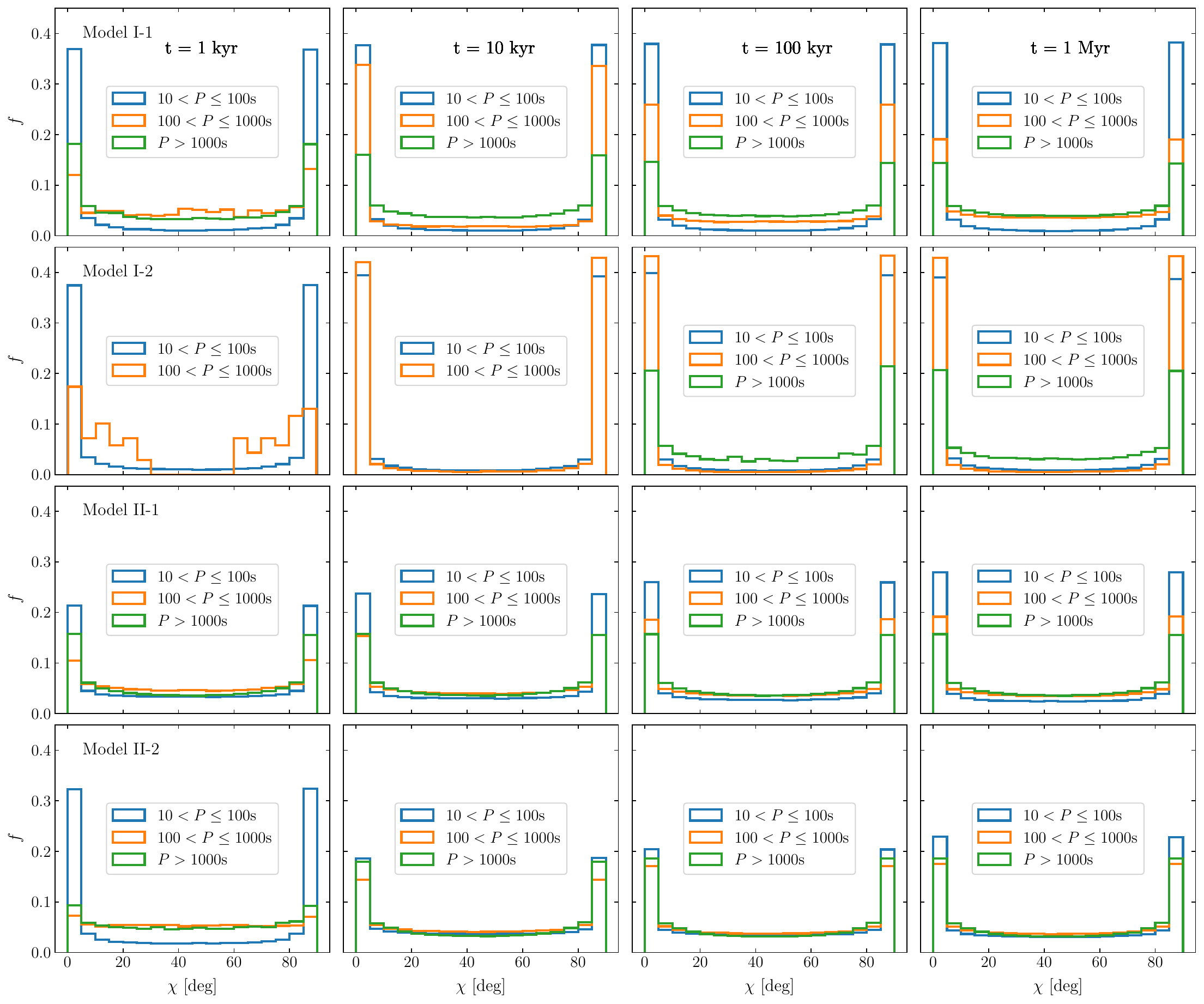}
	\caption{Evolution of the distribution of $\chi$ as a function of $P$ for the four models.}
	\label{fig:distri}
\end{figure*}

There are currently four ULPPs, i.e., PSR J0901 \citep{2022NatAs...6..828C}, GLEAM-X J1627 \citep{2022Natur.601..526H}, GPM J1839 \citep{hurley-walker_long-period_2023}, ASKAP J1935 \citep{2024NatAs.tmp..107C}, and one ultra long period X-ray source 1E 1613 \citep{2006Sci...313..814D}. The former have been detected in radio, indicating that they are in the ejector phase, while 1E 1613 has only been observed in X-ray band, probably in the propeller or weak accretor phase. In this section, we present  Monte-Carlo (MC) simulations to examine the feasibility of the models used in this paper to reproduce these ULPPs. We also explore the potential for detecting radio radiation from them by considering their expected radio fluxes and the evolution of the inclination angles.

\subsection{Monte-Carlo simulation for different evolutionary stages}\label{s:mc}

In the MC simulations, the initial distributions of $M_{\rm d0}$, $R_0$, $B_0$ and $P_0$ are assumed to follow log-uniform distribution, that is, $\log (M_{\rm d0}/M_{\odot})\in [-7,-1]$, $\log (R_{0}/R_{\rm s})\in [1,6]$, $\log B_{0}({\rm G})\in [13,16]$, and $\log P_{0}({\rm s})\in [-3,1]$, both $\alpha_0$ and $\chi_0$ are uniformly distributed between 0 and $\pi/2$. We evolve one million sources, and display the simulated results of $P$ and $\dot{P}$ at $t=10^3$, $10^4$, $10^5$, and $10^6$ yr in Figure~\ref{fig:mc} (from left to right). Different phases of the NSs are distinguished by the colored dots. The red stars represent the four observed ULPPs (PSR J0901, GLEAM-X J1627, GPM J1839, and ASKAP J1935), and the red line 1E 1613.

The first row of Figure~\ref{fig:mc} displays the results in model I-1. There are two populations of ejectors, with the shorter period one distributed in the lower left and the longer period one in the lower right. The former have only experienced spin-down by magnetic dipole radiation, which means that there has been no disk formed around the NS. The latter have been spun down by the fallback disks and subsequently exhibit as radio pulsars after the disks become inactive. The four ULPPs can be covered by the simulated distributions of ejectors, but it is difficult to reach the very long spin period of 1E 1613 during the propeller/accretor phase due to the relatively short duration of the NS-disk interaction. 

The second row demonstrates the results in model I-2. Most of the simulated sources are in the accretor and propeller phases because the fallback disks survive for a longer time than in model I-1. The NSs can be spun down to $>10^4$ s due to prolonged interaction with the disks, which may account for the spin period of 1E 1613, but it is hard to explain the four ULPPs.

The lower two rows present the results in  model II. In the third row the lifetime of the disk is so short that after $10^4$ yr no disk exists. Similar as in the first row, the simulation can reproduce the periods of the four ULPPs but cannot reproduce the extremely long period of 1E 1613.
In the bottom row, most of the NSs have experienced the interaction with the disk, which survive for $\lesssim 10^5$ yr. This model is able to account for the four ULPPs, with the longest attainable spin period reaching a few $10^4$ s. However, it falls slightly short of fully reproducing 1E 1613.

\subsection{The death line of radio pulsars}\label{s:death}
A NS in the ejector phase is not necessarily a radio pulsar.
The death line of radio pulsars is defined as the line in the $P-\dot{P}$ or $P-B$ diagram to distinguish between pulsars capable of sustaining pair production in their magnetospheres and those that cannot \citep{1975ApJ...196...51R}. Generally, the fiducial death line is derived under the assumption that the maximum acceleration potential across the polar cap region equals $10^{12}$ V, below which the radio emission is quenched. Moreover, there have been various models to define the criteria, including the ``death valley'' suggested by \citet{1993ApJ...402..264C} which depends on different configuration of the NS magnetic field \citep[e.g.,][]{Zhang2000,2002ApJ...565..482G,2015MNRAS.450.1990K}, and the modified death line with a fallback disk \citep{2023ApJ...943....3T}. However, it seems  challenging to explain the radio activities of ULPPs, especially for GPM J1839 \citep{2023RAA....23l5018T,2024ApJ...961..214R}, from their positions in the $P-\dot{P}$ plane. 

It should be mentioned that the death lines discussed above are usually based on the assumption of an orthogonal rotator in vacuum, while pulsars are oblique rotators surrounded by plasma. \citet{2006ApJ...643.1139C} derived a spin-down formula that takes into account the misalignment of the magnetic and rotation axes and the magnetospheric particle acceleration gaps. Based on this model, \citet{2012ApJ...757L..10T} proposed a modified death line depending on the magnetic inclination $\chi$, 
\begin{equation}
    \dot{P}_{\rm death}=5\times10^{-18}\left(\frac{P_{\rm death}}{1\rm s}\right)^3\left(\frac{V_{\rm gap}}{10^{12}\rm V}\right)^2\sin^2\chi. \label{eq:death1}
\end{equation}

In Figure~\ref{fig:death}, we plot the simulated NSs in the ejector phase in Model II-2 with $\chi$ in different intervals (from top to bottom), as well as the modified death lines  with $\chi=0.2^{\circ}$, $10^{\circ}$, $30^{\circ}$, and $90^{\circ}$ (the dashed lines). When $\chi>30^{\circ}$, all the four observed sources are below the death line, suggesting their incapability of radio emission. When taking a smaller $\chi$ (i.e., $10^{\circ}$), both PSR J0901 and GLEAM-X J1627 are above the death line, implying they might be capable of emitting in radio if possessing a relatively small $\chi$. In the more extreme case, if $\chi$ is less than $1^{\circ}$ ($0.2^{\circ}$), then ASKAP J1935 (GPM J1839) becomes above the death line\footnote{Alternatively, \citet{2000ApJ...531L.135Z} suggested that the inverse Compton scattering–induced space-charge–limited flow may sustain strong pair production in some long-period pulsars without introducing anomalous magnetic field configurations of the NSs, and thus allowing them to be radio-loud pulsars. The death line in this model for dipolar magnetic field configurations is
    $\log\dot{P}_{\rm death}=-\frac{3}{11}\log P_{\rm death}-15.36$, 
which is depicted as the gray solid lines in Figure~\ref{fig:death}. It is apparent that all of the simulated and observed ULPPs are well above this line, which may imply a large parameter space for the radio emission of ULPPs.}.

Obviously the ULPPs should not be aligned rotators. However, the above analyses suggest that the magnetic field configuration could be an important role in determining the radio properties of ULPPs.
We then examine the probability distribution of $\chi$ of the NSs at the ejector phase in different intervals of $P$, as shown in Figure~\ref{fig:distri}. It is clearly seen that the distribution of $\chi$ exhibits a U-shaped pattern in most cases, which means that a large fraction of ULPPs could be nearly aligned and orthogonal rotators. We find that the fraction $f_{\rm peak}$ of NSs with $\chi$ smaller than $5^{\circ}$ and larger than $85^{\circ}$ varies with both time and $P$. Taking the results of Model II-1 for example (the third row in Figure~\ref{fig:distri}): when $10\ {\rm s}<P\le 100\ \rm s$, $f_{\rm peak}$ increases from 0.42 at $t=1$ kyr to 0.56 at $t=1$ Myr; when $100\ {\rm s}<P\le 1000\ \rm s$, $f_{\rm peak}$ first increases rapidly within $\sim 100$ kyr from 0.2 to 0.38, and stabilizes around 0.4 eventually at $t=1$ Myr; when $P> 1000\ \rm s$, $f_{\rm peak}$ remains to be approximately 0.3. 

\subsection{The radio detectability}\label{s:flux}

To examine the detectability of ULPPs in radio, we simulate their flux distribution at 1.4 GHz in different models and compare the results with the observations. The pseudo-luminosity at 1.4 GHz of pulsars is approximated to be \citep{2006ApJ...643..332F}
\begin{equation}
    \log L_{\nu,1.4\rm GHz}=\log(L_0P_0^{\epsilon_{P}}\dot{P}_{15}^{\epsilon_{\dot{P}}})+L_{\rm corr}, \label{eq:lum}
\end{equation}
where $L_0=0.18\,\rm mJy\ kpc^2$, $\epsilon_{P}=-1.5$, and $\epsilon_{\dot{P}}=0.5$. We assume that $L_{\rm corr}$ follows a zero-centered normal distribution with the standard deviation $\sigma_{L_{\rm corr}}=0.8$.

To calculate the radio flux of the NSs at 1.4 GHz ($F_{\nu,1.4\rm GHz}=L_{\nu,1.4\rm GHz}/d^2$) we need to know their distribution in the the Milky Way and their distance $d$ to the Earth. We assume that the NS distribution follows the stellar distribution in the Milky Way \cite[e.g.,][]{Yamaguchi+2018}, and employ the number density ${\mathrm d N}/{\mathrm d l\,\mathrm d b\,\mathrm d d}$ as a function of the distance from the Galactic center $r=\sqrt{r_0^2+d^2\cos^2b-2d r_0\cos b\cos l}$ in the Galactic plane and the distance perpendicular to the Galactic plane $z=d\sin b$ \citep{Bahcall+1980}, i.e.,
\begin{equation}\label{eq:distantDistribution}
	\frac{\mathrm d N}{\mathrm d l~\mathrm d b ~\mathrm d d}\propto d^2\cos b \exp{\left(-\frac{z}{h_z}-\frac{r-r_0}{h_r}\right)},
\end{equation}
where $r_0=8.5~{\rm kpc}$ is the distance from the Galactic center to the Sun, $b$ and $l$ are respectively the Galactic latitude and longitude, and $(h_z,h_r)=(0.25~{\rm kpc},\,3.5~{\rm kpc})$ represent the scale lengths for the exponential stellar distributions perpendicular and parallel to the Galactic plane, respectively.

\begin{figure*}[t]
	\centering	\includegraphics[width=\textwidth]{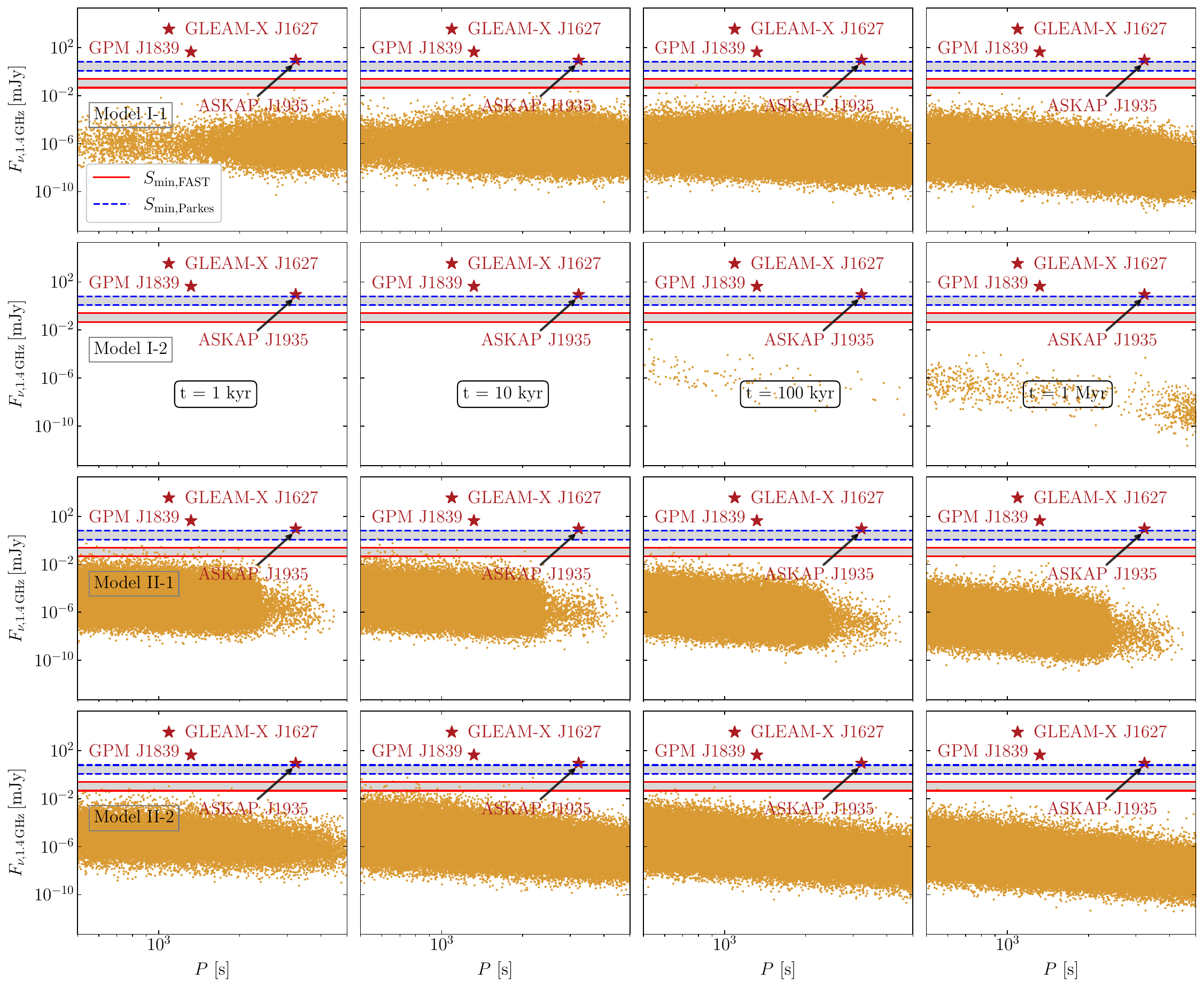}
	\caption{The simulated distribution of flux at 1.4GHz $F_{\nu,1.4\rm GHz}$ of ULPPs larger than 500s. The red stars represent the observed flux of GLEAM-X J1627, GPM J1839 and ASKAP J1935 derived by their radio spectrum. The red solid and blue dashed lines denote the sensitivity limits of the FAST and Parkes telescopes, respectively.}
	\label{fig:flux1}
\end{figure*}

The simulated radio fluxes of ULPPs at different ages are shown in Figure~\ref{fig:flux1}, as a function of $P$. The detection limit for radio transient bursts is estimated as follows \citep[e.g.,][]{2003ApJ...596..982M}:

\begin{equation}
    S_{\rm min}={SN}\times \frac{\beta S_{\rm sys}}{\sqrt{N_{\rm p}\Delta \nu W}},
\end{equation}
where $SN=10$ is the required signal-to-noise ratio, $\beta$ is a correction factor to account for 8-bit data digitization, $S_{\rm sys}$ is the system equivalent flux densities, $N_{\rm p}$ is the number of polarizations, $\Delta \nu$ is the bandwidth, and $W$ is the observed width of the pulse. 
The red solid and blue dashed lines in Figure~\ref{fig:flux1} represent the detection limits for the Five-hundred-meter Aperture Spherical radio Telescope \citep[FAST,][]{Nan+2008,Nan+2011,Jiang+2020}\footnote{\url{https://fast.bao.ac.cn}} and the 64-metre Parkes radio telescope (Parkes)\footnote{\url{https://www.parkes.atnf.csiro.au}}, respectively. We take $S_{\rm sys}\simeq 1.5~{\rm Jy}$, $\beta=1.5$, $N_{\rm p}=2$, $\Delta \nu=400~{\rm MHz}$, and $W=10-300~{\rm s}$ \citep[for the typical pulse widths of known ULPPs,][]{2022Natur.601..526H,hurley-walker_long-period_2023,2024NatAs.tmp..107C} for the L-band receiver of FAST \citep{Jiang+2020}, and $S_{\rm sys}\simeq38~{\rm Jy}$, $\beta=1.5$, $N_{\rm p}=2$, and $\Delta \nu \simeq 400~{\rm MHz}$ for the ultra-wide bandwidth receiver's subbands 3, 4, and 5 of Parkes \citep{Hobbs+2020}.
The three red stars depict the observed values of GLEAM-X J1627, GPM J1839 and ASKAP J1935 derived by the fluxes at lower frequency and conjectural radio spectrum, probably corresponding to an upper limit of the flux at 1.4 GHz. It is seen that the simulated fluxes are much lower than the inferred ones from the observations and below the detection limit of both FAST and Parkes, indicating either there are abnormal radio radiation mechanisms for ULPPs, or the derived radio energy spectrum may be not suitable for the higher frequency bands. 

\begin{figure}[t]
	\centering	\includegraphics[width=.5\textwidth]{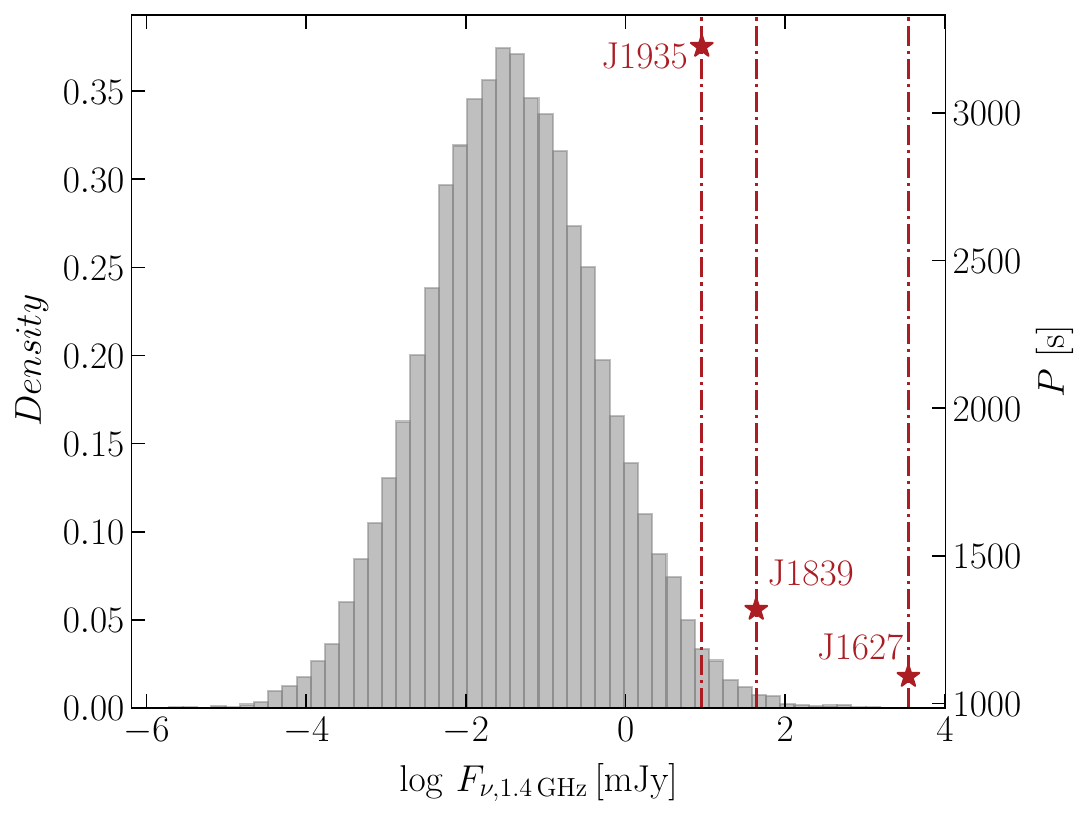}
	\caption{The density histogram of the 1.4 GHz flux distribution of radio pulsars. The right axis indicates the spin periods of the three ULPPs. The red lines represent the observed fluxes of GLEAM-X J1627, GPM J1839 and ASKAP J1935 derived from their radio spectrum.}
	\label{fig:flux2}
\end{figure}

\begin{figure}[t]
	\centering	\includegraphics[width=.5\textwidth]{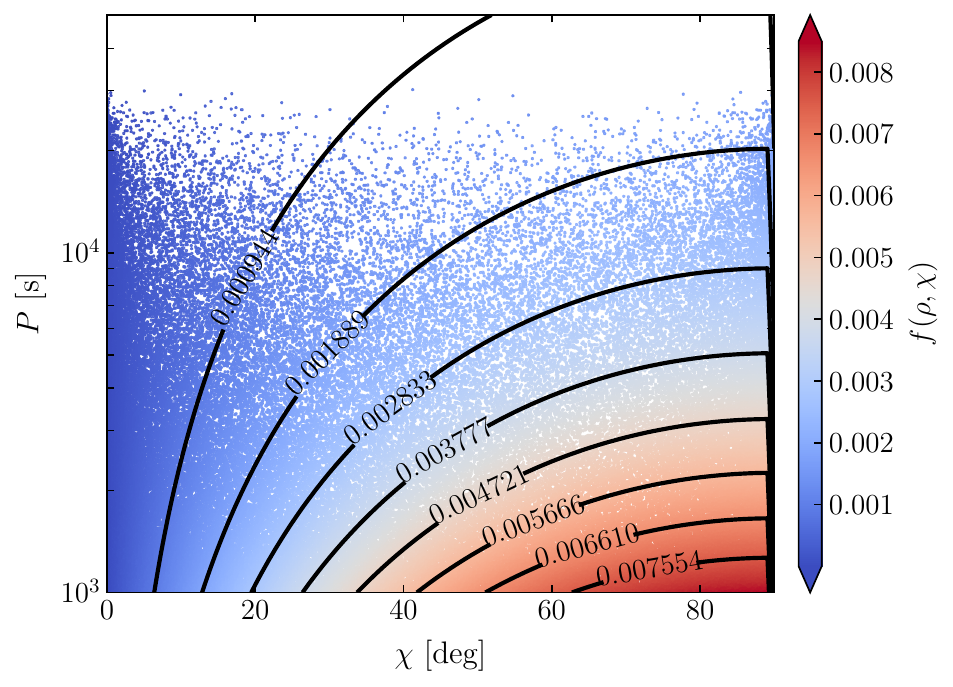}
	\caption{The distribution of the beam fraction in the $P-\chi$ space for Model II-2 at the age of 10 kyr.}
	\label{fig:beam}
\end{figure}

Alternatively, \citet{2014ApJ...784...59S} showed that the observed radio luminosities of pulsars weakly depend on the spin period and period evolution, but follow a log-normal distribution. We thus adopt the following log-normal luminosity distribution
\begin{equation}
    \log L_{\nu,1.4\, \rm GHz}\sim N(0.5,1.0),
\end{equation}
and calculate the 1.4 GHz radio flux distribution. Figure~\ref{fig:flux2} demonstrates the probability density of $\log F_{\nu,1.4\, \rm GHz}$. The left y-axis represents the distribution density, and the right y-axis is set to be the spin period for comparison with the three ULPPs. Similar as in  Figure~\ref{fig:flux1}, the derived fluxes of the ULPPs are still higher than expectation, especially for GLEAM-X J1627.

Moreover, the radio detectability of pulsars depends not only on the radio flux but also on the geometry of the radio beam and whether it intersects the line of sight. The beam fraction $f$ of a pulsar is given by \citep{1998MNRAS.298..625T}
\begin{align}
f(\rho,\chi)=
\begin{cases}
 &2\sin\chi\sin\rho,\ \mathrm{if} \ \chi>\rho\ \mathrm{and}\ \chi+\rho<\frac{\pi}{2};\\
 &\cos(\chi-\rho),\ \mathrm{if} \ \chi>\rho\ \mathrm{and}\ \chi+\rho>\frac{\pi}{2};\\
 &1-\cos(\chi+\rho),\ \mathrm{if} \ \chi<\rho\ \mathrm{and}\ \chi+\rho<\frac{\pi}{2};\\
 &1,\ \mathrm{if} \ \chi<\rho\ \mathrm{and}\ \chi+\rho>\frac{\pi}{2},\\
\end{cases}
\end{align}
where $\rho$ is the beam radius for a circular beam. A lower $f$ value indicates a lower detection probability. According to Figure~\ref{fig:distri}, in the samples that can spin down to the periods exceeding 1000 s, 30 percent or more of the systems have $\chi$ values less than $5^{\circ}$ or greater than $85^{\circ}$. {\em If the beam radii of ULPPs follow a relation with $P$ similar to those of normal pulsars} \citep{1994PhDT.......164G}, i.e. $\rho(P)=7.7^{\circ}P^{-1/2}$, we can derive the $f$ distribution in the $\chi-P$ space. Figure~\ref{fig:beam} shows an example for Model II-2 at $t=10$ kyr. For systems with $\chi>85^{\circ}$, $f$ ranges between 0.001 and 0.008; while for systems with $\chi<5^{\circ}$, $f$ is generally below 0.001. The very small values of $f$ indicate that most ULPPs might be undetectable.

\section{Conclusions}\label{s:conclusion}

In this paper, we investigate under what conditions strong-magnetized NSs born with fallback disks could evolve to be ULPPs. We have tracked the (P, $\alpha$ and $\chi$)  evolution of the NSs, considering different parameters related to both the disk and the NSs. In particular, we take the evolution of disk into account and find that the lifetime of an active fallback disk has significant impact on the formation and radio observational properties of ULPPs. We consider different neutralization temperatures in the outer disk  \citep{2001ApJ...559.1032M,2001ApJ...554.1245A}, and carry out Monte-Carlo simulations of the NS evolution. Our results reveal that a fallback disk is necessary to spin down the NS to very long period. However, its lifetime needs to be limited so that the NS is able to evolve out of the disk-fed phase to the ejector phase. This critical factor has be neglected in most of the previous works on the similar topic, which only focused on the spin period evolution without accounting for its evolutionary state. In this regard, Model II-2, which assumes that the NS-disk interaction terminates once the effective temperature in the outer disk radius reaches 300 K, seems to be the most adequate, although it remains slightly insufficient in fully reproducing 1E 1613.

However, some extra mechanisms seem to be required to account for radio emission of ULPPs because all of them are well below the traditional death line. We notice that the death lines with small magnetic oblique angle $\chi$ could potentially present an alternative interpretation for their radiation characteristic, and the distribution of $\chi$ of simulated ULPPs is indeed clustered around $0^{\circ}$ and $90^{\circ}$, which means that a sizable fraction of ULPPs could be nearly aligned or orthogonal rotators. If the beam radii of ULPPs follow the same relationship as those of normal pulsars, the beam fraction would be very small, making most ULPPs undetectable.
Moreover, the simulated distribution of the 1.4 GHz fluxes of ULPPs are significantly lower than the derived observed values of GLEAM-X J1627 and GPM J1839. This indicates that our understanding of the radiation mechanisms of ULPPs is far from being complete.

We caution that our analysis on the radio emission of ULPPs is based on the traditional rotation-powered pulsar models \citep{1975ApJ...196...51R}, which has been strongly challenged by the observed period derivative upper limits and bright radio emission of ULPPs. Recently, \citet{2024MNRAS.533.2133C} suggested that plastic motion and/or thermoelectrically driven twists in the crusts of slowly rotating ($P\gtrsim 100$ s), highly magnetized ($B\gtrsim 10^{14}$ G) NSs could impart mild local magnetospheric twists. The initiated cascades via resonant inverse-Compton scattered photons or curvature radiation may produce broad-band coherent radio emission. The derived radio active zones in the $P-\dot{P}$ diagram seem to be compatible with observed ULPPs. However, this model critically depends on the plastic flow velocity, which is currently poorly constrained.

Finally, although we focus on the evolution of NSs with a supernova fallback disk, the results are aslo relevant to NSs born from the merger of double white dwarfs, which are also likely surrounded by a fossil disk made of the disrupted white dwarf material \citep{1990ApJ...348..647B,2004A&A...413..257G,2006ApJ...644.1063D,2009A&A...500.1193L,2012ApJ...746...62R,2013ApJ...767..164Z,2014MNRAS.438...14D,2018ApJ...857..134B}. This type of NSs may be closely related to young pulsars \citep{1996ApJ...460L..41L,2011ApJ...742...51B} and fast radio bursts \citep{2021ApJ...910L..18B,2022Natur.602..585K} discovered in globular clusters. This topic was recently discussed in the literature \citep{2021ApJ...917L..11K,2022MNRAS.510.1867L,2023MNRAS.525L..22K}, but the effect of the remnant disk has not been explored. Figure 3 suggests that it is difficult to form millisecond pulsars through merger-induced collapse of white dwarfs.

\section*{acknowledgments}
We are grateful to the referee for helpful comments. This work was supported by the National Key Research and Development Program of China (2021YFA0718500), the Natural Science Foundation of China under grant No. 12041301, 12121003 and 123B2045.

\section*{Data Availability}
The simulated MC data underlying this
article are available at \dataset[zenodo.13788218]{https://doi.org/10.5281/zenodo.13788218}. The other data and codes underlying this article will be shared on reasonable request to the authors.

\bibliography{YL23}{}
\bibliographystyle{aasjournal}

\end{document}